\documentclass[aps,epsfig,rotate,showpacs]{revtex4}  

\begin{document}                
%
%
\title{When is Quantum Decoherence Dynamics Classical?}
\author{Jiangbin Gong and Paul Brumer}
\affiliation{ Chemical Physics Theory Group, 
Department of Chemistry,\\
University of Toronto,Toronto, Canada  M5S 3H6}
\date{\today}

\begin{abstract}              
A direct classical analog of quantum decoherence is introduced. 
Similarities and differences between decoherence dynamics examined quantum
mechanically and classically are exposed via a second-order perturbative
treatment and via a strong decoherence theory, showing a strong dependence
on the nature of the system-environment coupling. For example, 
for the traditionally assumed linear coupling, the classical and
quantum results are shown to be in exact agreement.
\end{abstract}


\maketitle

Decoherence is the loss of quantum coherence due to system-bath coupling.
There has been considerable theoretical and experimental work
demonstrating that quantum-classical correspondence (QCC) can be induced
by decoherence \cite{zurekreview,gong,raizen}. By contrast, little work has
been done on examining the correspondence between classical and quantum
descriptions of the time evolution of decoherence itself, i.e. decoherence
dynamics. In this Letter we show that (a) one can introduce a direct
classical analog of quantum decoherence, and (b) by examining the dynamics
of decoherence classically one gains new insights into both the dynamics
of decoherence described quantum mechanically and into the conditions for
QCC of the dynamics of decoherence. For example, we show that the extent
of QCC depends strongly on the nature of the system-bath coupling and far
less upon $\hbar$ than expected, that results assumed to be quantum
mechanical can be obtained classically and that nonlinear system-bath
coupling can cause nonclassical decoherence dynamics even for macroscopic
systems.

The formal Liouville-based theory of QCC in an isolated
system \cite{wilkie} makes clear that there is a strict analogy
between quantum and classical dynamics in phase space. As in the quantum
case, the classical Liouville dynamics of a closed system is unitary and
we expect the reduced classical Liouville dynamics of a system coupled to
a bath to be nonunitary. We therefore suspect that, due to bath's
coarse-graining effects, the reduced dynamics of the system propagated
classically will show decoherence dynamics that is, qualitatively,
parallel to that seen in quantum dynamics insofar as the loss of phase
information, entropy production, etc.

Here we quantitatively compare the dynamics of decoherence that is induced
quantum mechanically to that induced classically. This is done by
analytically examining the dynamics of an initial quantum state, in a
system coupled to a bath, that is propagated either quantum mechanically
or classically. The observed (rather remarkable) similarities and
differences between the classical and quantum decoherence dynamics should
be of considerable interest to a variety of modern fields such as quantum
information processing \cite{qcbook} and quantum control of atomic and
molecular processes \cite{brumerbook}. Further, this analysis 
offers new insights into both decoherence and QCC, is
relevant to semiclassical descriptions of decoherence dynamics
\cite{miller}, and has motivated purely classical descriptions of
dynamics-induced intrinsic decoherence, with preliminary 
computations \cite{gong2} that support the analytic results presented here.

We begin by introducing classical analogs of some
representation-independent and representation-dependent measures of
decoherence.  A second-order perturbative treatment is then used to
examine QCC in early-time decoherence dynamics. Subsequently, we introduce
a classical theory of strong decoherence that allows us to go beyond
perturbation theory; the results are then compared with corresponding
quantum theory.

Consider a system with a time-independent Hamiltonian
$H^{s}=P^{2}/(2m)+V(Q)$, where $(Q,P)$ are conjugate position and momentum
variables, coupled to $N$ independent harmonic bath modes described by the
Hamiltonians
$H^{b}_{j}=p_{j}^{2}/(2m_{j})+m_{j}\omega_{j}^{2}q_{j}^{2}/2$, where
$\{p_{j}, q_{j}\}$ ($j=1,2,\cdots, N$) are bath-mode phase space
variables. The system-bath coupling potential is assumed to be
$V^{sb}=\sum_{j=1}^{N}C_{j}f(Q)q_{j}$ so that the total Hamiltonian is
given by $H=H^{s}+\sum_{j=1}^{N} [H^{b}_{j}+C_{j}f(Q)q_{j}]$. The
phase space distribution function evolved classically
and the quantum Wigner function for the entire
phase space are represented by $\rho^{c}[Q,P,\{q_{j},p_{j}\},t]$ and
$\rho^{W}[Q,P,\{q_{j},p_{j}\},t]$, respectively. Their time evolution is
given by $\partial \rho^{c}/\partial t=\{H,\rho^{c}\}_{P}$ and $\partial
\rho^{W}/\partial t=\{H,\rho^{W}\}_{M}$, where $\{\cdot\}_{P}$ denotes
classical Poisson bracket and $\{\cdot\}_{M}$ denotes quantum Moyal
bracket. Further, we define classical and quantum reduced distribution
functions $\tilde{\rho}_{c}(Q,P,t)\equiv\int \rho^{c}d\Gamma^{N}_{b}$ and
$\tilde{\rho}_{W}(Q,P,t)\equiv\int \rho^{W}d\Gamma^{N}_{b}$, where
$d\Gamma^{N}_{b}\equiv\prod_{j=1}^{N}dq_{j}dp_{j}$. Our interest is in the
correspondence between the classical and quantum decoherence dynamics of
initial states that can be either classical (i.e. positive Wigner density
everywhere) or nonclassical \cite{wilkie} (e.g. displaying regions of
negative $\rho^{W}[Q,P,\{q_{j},p_{j}\},t]$).

Consider two measures of decoherence in each of classical and
quantum mechanics. One widely-used and representation-independent
measure is the linear entropy $S_{q}\equiv
1-Tr(\hat{\tilde{\rho}}^{2})$ \cite{linear-entropy}, where
$\hat{\tilde{\rho}}$ is the reduced density operator of the
system. An increase in $S_{q}$ causes $1/(1-S_{q})$ to increase,
corresponding to an increasing number of incoherently populated
orthogonal quantum states. Since
$S_{q}=1-2\pi\hbar\int\tilde{\rho}_{W}^{2}(Q,P,t)d\Gamma_{s}$,
where $d\Gamma_{s}\equiv dQdP$, this entropy has a natural
classical analog (denoted $S_{c}$) obtained by replacing
$\tilde{\rho}_{W}$ with $\tilde{\rho}_{c}$. That is, $S_{c}\equiv
1-2\pi\hbar\int\tilde{\rho}_{c}^{2}(Q,P,t)d\Gamma_{s}$. A more
detailed, but representation-dependent description of decoherence
is the decay of off-diagonal density matrix elements such as
$\langle Q_{1}|\hat{\tilde{\rho}}(t)|Q_{2}\rangle$. Significantly,
we discover that the classical analog of these matrix elements can also be
constructed. Specifically, noting that $\langle
Q_{1}|\hat{\tilde{\rho}}(t)|Q_{2}\rangle=\int dP
\tilde{\rho}_{W}(\overline{Q},P ,t)\exp\left[ i\Delta Q
P/\hbar\right]$, where $\overline{Q}\equiv (Q_{1}+Q_{2})/2$ and
$\Delta Q=Q_{1}-Q_{2}$, we define the classical analog (denoted
$\tilde{\rho}_{c}(Q_{1},Q_{2},t)$) of $\langle
Q_{1}|\hat{\tilde{\rho}}(t)|Q_{2}\rangle$ as the Fourier
transformed classical distribution function, i.e.,
$\tilde{\rho}_{c}(Q_{1},Q_{2},t)\equiv \int dP
\tilde{\rho}_{c}(\overline{Q} ,P,t)\exp\left[i\Delta
QP/\hbar\right]$. This approach can be readily extended to the
momentum representation.

Perturbative treatments have proved to very useful in understanding
decoherence dynamics \cite{kim,perturbation-approaches}. Here, to examine
classical vs. quantum decoherence dynamics at short times, a regime of
great interest in the control of decoherence, we consider a second-order
expansion with respect to time variable $t$ for both $S_{c}$ and $S_{q}$,
i.e., $S_{c}(t) = S_{c}(0)+t/\tau_{c,1}+t^{2}/\tau_{c,2}^{2}+\cdots$, and
$S_{q}(t) = S_{q}(0)+t/\tau_{q,1}+t^{2}/\tau_{q,2}^{2}+\cdots$. Using the
definitions of Poisson and Moyal brackets and assuming that the initial
distribution function is decorrelated with initial bath statistics, we
obtain
\begin{equation}
\frac{1}{\tau_{c,1}}=\frac{1}{\tau_{q,1}}=0,
\label{firstorder}
\end{equation}
\begin{equation}
\frac{1}{\tau_{c,2}^{2}}=\frac{C_{b}}{\hbar}\int dQ_{1}dQ_{2}
|\tilde{\rho}_{c}(Q_{1},Q_{2},0)|^{2}\Delta Q^{2}
\left[\frac{df(\overline{Q})}{d\overline{Q}}\right]^{2},
\label{cla-2nd}
\end{equation}
and
\begin{eqnarray}
\frac{1}{\tau_{q,2}^{2}}=\frac{C_{b}}{\hbar}
\int dQ_{1}dQ_{2}|\langle Q_{1}|\hat{\tilde{\rho}}(0)|Q_{2}\rangle|^{2}\Delta Q^{2}
\left[\frac{\Delta f(\overline{Q})}{\Delta Q}\right]^{2},
\label{quan-2nd}
\end{eqnarray}
where $C_{b}=
\sum_{j=1}^{N}C_{j}^{2}\coth(\beta\hbar\omega_{j}/2)/(2m_{j}\omega_{j})$,
$\Delta f(\overline{Q})\equiv f(\overline{Q}+\Delta
Q/2)-f(\overline{Q}-\Delta Q/2)$, and $\beta$ is the Boltzmann factor.
Note that the factor $\hbar$ appearing in the classical result [Eq.
\ref{cla-2nd})] is just due to the definitions of $S_{c}$
and $\tilde{\rho}_{c}(Q_{1},Q_{2}, 0)$, and that the initial
variances of the bath variables $q_{j}$ have been evaluated using quantum
statistics to ensure the same initial quantum state for
the ensuing classical and quantum dynamics. Note also that the decoherence
time scale indicated in the easily-derived and simple quantum result of
Eq. (\ref{quan-2nd}) is consistent with, but is more transparent than, a
previous perturbation result (Eq. (5.6) in Ref. \cite{hu}) obtained using
a sophisticated influence functional approach.

Equation (\ref{firstorder}) shows that zero first-order decoherence rate
i.e., $1/\tau_{q,1}$=0, has a strict classical analog. More interestingly,
Eqs. (\ref{cla-2nd}) and (\ref{quan-2nd}) show that, for the same fixed
initial distribution function, the ratio of $1/\tau_{q,2}^{2}$ to
$1/\tau_{c,2}^{2}$ is $\hbar$-independent. As seen from Eqs.
(\ref{cla-2nd}) and (\ref{quan-2nd}),
$(1/\tau_{q,2}^{2}-1/\tau_{c,2}^{2})$ arises from the difference between
the derivative $df/dQ$ and the finite-difference function $\Delta f/\Delta
Q$, weighted by $\Delta Q^{2}$ and the initial state. As a result: (1) For
any given $f(Q)$, as long as $\langle
Q_{1}|\hat{\tilde{\rho}}(0)|Q_{2}\rangle$ decays fast enough with $|\Delta
Q|$ such that $\Delta f/\Delta Q \approx df/dQ$, there would be excellent
QCC in early-time decoherence dynamics. The smaller the $\hbar$, the more
rigorous is this requirement. (2) If $f(Q)$ depends only linearly or
quadratically upon the coupling coordinate $Q$, then
$(1/\tau_{q,2}^{2}-1/\tau_{c,2}^{2})=0$ for any initial state.
Significantly then, in all traditional decoherence models
\cite{linear-models} where $f(Q)=Q$ is assumed, there exists perfect QCC
in early decoherence dynamics, regardless of $\hbar$, and irrespective of
the system potential $V(Q)$ \cite{footnote1}. 
Indeed, in the case of
$f(Q)=Q$ Eq. (\ref{cla-2nd}) reduces to an important result, previously
obtained quantum mechanically \cite{kim}:
\begin{eqnarray}
\frac{1}{\tau_{c,2}^{2}}=\frac{1}{\tau_{q,2}^{2}}=
2\frac{\delta^{2}Q}{\hbar} \sum_{j=1}^{N}
\frac{C_{j}^{2}}{2m_{j}\omega_{j}}\coth(\frac{\beta\hbar\omega_{j}}{2}).
\end{eqnarray}
where the initial state of the system is assumed to be pure, with the
initial variance in $Q$ given by $\delta^{2} Q$.
(3) For nonlinear $f(Q)$ where $\Delta f/\Delta Q\ne df/dQ$ over the range of
the initial state, QCC can be very poor.

The second-order perturbative treatment is most reliable at short
times and for weak decoherence. The results are particularly
significant for studies of decoherence control where early-time
dynamics of weak decoherence is important. In these circumstances
it is useful to understand the extent to which (quantum)
decoherence is equivalent to classical entropy production, i.e. to
increasing $S_c(t)$. In particular, if there exists good
correspondence between classical and quantum decoherence dynamics,
then the essence of decoherence control is equivalent to the
suppression of classical entropy production, and various classical
tools may be considered to achieve decoherence control. If not,
then fully quantum tools are required.

As an example, consider decoherence for an initial superposition state of
two well-separated and strongly localized Gaussian wavepackets located at
$Q_{a}=\overline{Q}_{ab}-\Delta Q_{ab}/2$ and
$Q_{b}=\overline{Q}_{ab}+\Delta Q_{ab}/2$ with $\overline{Q}_{ab}=0$. For
this initial state, $1/\tau_{c,2}^{2} \sim (C_{b}/\hbar) \Delta
Q_{ab}^{2}\left [df(\overline{Q}_{ab})/d\overline{Q}_{ab}\right ]^{2}$,
and $ 1/\tau_{q,2}^{2} \sim (C_{b}/\hbar)\Delta Q_{ab}^{2}\left [\Delta
f(\overline{Q}_{ab})/\Delta Q_{ab}\right]^{2}$. Then in a cubic
decoherence model, for example, where $f(Q)=Q^{3}$, one would obtain
$1/\tau_{c,2}^{2}\sim 0$ since
$df(\overline{Q}_{ab})/d\overline{Q}_{ab}=0$. However, here
$1/\tau_{q,2}^{2}>>1/\tau_{c,2}^{2}$, i.e. there is appreciable {\it
decoherence without classical entropy production}. By contrast, in another
nonlinear decoherence model where $f(Q)=\sin(2\pi Q/\Delta Q_{ab}+\pi/4)$,
$1/\tau_{q,2}^{2}\sim 0$ since $f(Q_{a})=f(Q_{b})$. Here, however,
$1/\tau_{c,2}^{2}>>1/\tau_{q,2}^{2}$, i.e., the system is {\it
decoherence-free but with substantial classical entropy production}. Since
we find that the ratio of $\tau_{q,2}^{2}$ to $\tau_{c,2}^{2}$ in
early-time decoherence dynamics is independent of $\hbar$ for fixed
initial state, these two examples lead to a rather counter-intuitive
result: given a macroscopic object which is initially in a superposition
state of two distinguishable states and is nonlinearly coupled with an
environment, classical dynamics could totally fail to predict its initial
entropy production or its decoherence rate. Indeed, Eqs. (\ref{cla-2nd})
and (\ref{quan-2nd}) suggest that, as long as
$df(\overline{Q})/d\overline{Q}\ne 0$ and $|f(Q)|$ is bounded, then
$1/\tau_{q,2}^{2}$ saturates with increasing $\Delta Q_{ab}$, whereas
$1/\tau_{c,2}^{2}$ does not.  Thus, one can conclude
that decoherence dynamics must be quantum and that the system-environment 
coupling must be nonlinear 
if the saturation behavior of early-time
decoherence rates is  observed experimentally\cite{saturation-note}. Further, 
it is clear that in the limit of large $\Delta Q_{ab}$,
classical decoherence dynamics in the general case of nonlinear
system-environment coupling predicts much faster decoherence than does
quantum decoherence dynamics.  This leads to the rather surprising
inference that initial superposition states of well-separated wavepackets
would be more susceptible to nonlinear system-environment coupling
if they are propagated by classical dynamics than by quantum
mechanics.

To go beyond the perturbation results we now consider a strong
decoherence model in which decoherence is assumed to be much faster than
the system dynamics, so that $H^{s}$ can be set to zero \cite{zurekreview}.
We consider both the ``off-diagonal elements"
$\tilde{\rho}_{c}(Q_{1},Q_{2},t)$ as well as the entropy $S_c(t)$ and
compare them to the quantum results.

In this case
the classical Liouville dynamics gives
\begin{eqnarray}
& & \frac{\partial F_{c}[\overline{Q},\Delta Q,\{q_{j},p_{j}\},t]}{
\partial t}
 =  \sum_{k=1}^{N}
\frac{\partial H_{k}^{b}}{\partial q_{k}}
\frac{\partial F_{c}[\overline{Q},\Delta Q,\{q_{j},p_{j}\},t]}{\partial
p_{k}}
 \nonumber \\
 && -\sum_{k=1}^{N}\frac{\partial H_{k}^{b}}{\partial p_{k}}
 \frac{\partial F_{c}[\overline{Q},\Delta Q,\{q_{j},p_{j}\},t]}{\partial
 q_{k}}
 + \sum_{k=1}^{N}C_{k}f(\overline{Q})\frac{\partial
 F_{c}[\overline{Q},\Delta Q,\{q_{j},p_{j}\},t]}{\partial p_{k}} \nonumber \\
 &&-
 \frac{i}{\hbar}\Delta Q
 \sum_{k=1}^{N}C_{k}\frac{df(\overline{Q})}{d\overline{Q}}q_{k}
 F_{c}[\overline{Q},\Delta Q,\{q_{j},p_{j}\},t],
 \label{liouville2}
 \end{eqnarray}
where $F_{c}(\overline{Q},\Delta Q,\{q_{j},p_{j}\},t)\equiv \int dP
\exp[i\Delta Q P/\hbar] \rho_{c}[\overline{Q},P,\{q_{j},p_{j}\},t]$.
Since $\dot{Q}=0$ due to $H_{s}=0$,  and $\Delta Q$ is a
time-independent parameter introduced in the Fourier
transformation, Eq. (\ref{liouville2}) leads to
\begin{eqnarray}
 && \frac{d F_{c}[\overline{Q},\Delta Q,\{q_{j}(t),p_{j}(t)\},t]}{dt}
 = \frac{\partial F_{c}[\overline{Q},\Delta Q,\{q_{j}(t),p_{j}(t)\},t]}{
\partial t} \nonumber \\
& & + \sum_{k=1}^{N}
\frac{\partial F_{c}[\overline{Q},\Delta Q,\{q_{k}(t),p_{k}(t)\},t]}{
\partial q_{k}(t)}\dot{q}_{k}(t)
+\sum_{k=1}^{N}\frac{\partial F_{c}[\overline{Q},\Delta Q,\{q_{k}(t),p_{k}(t)\},t]}{
\partial p_{k}(t)}\dot{p}_{k}(t)
\nonumber \\
&&= -
\frac{i}{\hbar}\Delta Q
\sum_{k=1}^{N}C_{k}q_{k}(t)\frac{df(\overline{Q})}{d\overline{Q}}
F_{c}[\overline{Q},\Delta Q,\{q_{j}(t),p_{j}(t)\},t],
\label{liouville3}
\end{eqnarray}
where $\{q_{j}(t),p_{j}(t)\}$ satisfy
$\dot{q}_{j}(t) =\partial H_{j}^{b}/\partial p_{j}(t)$ and
$\dot{p}_{j}(t) = -\partial H_{j}^{b}/\partial q_{j}(t)-C_{j}f(\overline{Q})$,
of which the solution is
\begin{eqnarray}
q_{j}(t)= \frac{C_{j}f(\overline{Q})}{m_{j}
\omega_{j}^{2}}[\cos(\omega_{j}t)-1]
+q_{j}(0)\cos(\omega_{j}t)
+ \frac{p_{j}(0)}{m_{j}\omega_{j}}\sin(\omega_{j}t),
\label{traj3}
\end{eqnarray}
and $p_{j}(t)=m_{j}\dot{q}_{j}(t)$. Analytically integrating
Eq. (\ref{liouville3}), and
using $d\Gamma_{b}^{N}(t)=d\Gamma_{b}^{N}(0)$ and
$\tilde{\rho}_{c}(Q_{1},Q_{2},t)=
\int d\Gamma_{b}^{N}(t)F_{c}[\overline{Q},\Delta Q,\{q_{j}(t),p_{j}(t)\},t]$,
we have
\begin{eqnarray}
& & \tilde{\rho}_{c}(Q_{1},Q_{2},t)
=\int d\Gamma_{b}^{N}(0)F_{c}[\overline{Q},\Delta Q,\{q_{j}(0),p_{j}(0)\},0]
\nonumber \\
&& \times \exp[-\frac{i}{\hbar}\int^{t}_{0} dt \Delta Q
\sum_{k=1}^{N}C_{k}\frac{df(\overline{Q})}{d \overline{Q}}q_{k}(t)],
\label{cla-de1}
\end{eqnarray}
Substituting Eq. (\ref{traj3}) into Eq. (\ref{cla-de1}),
using the initial quantum state of the bath that is initially
uncorrelated with the system, and assuming that the equilibrium state
of the bath is maintained, we obtain
\begin{eqnarray}
  \frac{\tilde{\rho}_{c}(Q_{1},Q_{2},t)}{
 \tilde{\rho}_{c}(Q_{1},Q_{2},0)}=\exp
  \left[i\phi_{c}(t)
     -(\Delta Q)^{2}\left
   (\frac{df(\overline{Q})}{d\overline{Q}}
   \right)^{2}
   B_{2}(t)\right],
   \label{cresult2}
   \end{eqnarray}
where $\phi_{c}(t)\equiv
(\Delta Q)f(\overline{Q})[df(\overline{Q})/d\overline{Q}]B_{1}(t)/\hbar$, with
$B_{1}(t)=\sum_{j=1}^{N}
C_{j}^{2}[t-
\sin(\omega_{j}t)/\omega_{j}]
/(m_{j}\omega_{j}^{2})$,
and
$B_{2}(t)=
 \sum_{j=1}^{N}
  C_{j}^{2}\coth(\beta \hbar \omega_{j}/2)[1-\cos(\omega_{j}t)]/(2m_{j}\hbar
   \omega_{j}^{3})$.
Interestingly, the classical result [Eq. (\ref{cresult2})] displays
two dynamical aspects of 
$\tilde{\rho}_{c}(Q_{1},Q_{2},t)$, i.e., coherent dynamics of its
phase $\phi_{c}(t)$, and incoherent decay due to bath
statistics. The classical linear entropy $S_{c}(t)$ can  then be
obtained from Eq. (\ref{cresult2}) as
\begin{eqnarray}
S_{c}(t)=1-\int dQ_{1}dQ_{2}
|\tilde{\rho}_{c}(Q_{1},Q_{2},0)|^{2}\exp\left[-2(\Delta Q)^{2}
\left(\frac{df(\overline{Q})}{d\overline{Q}}\right)^{2}B_{2}(t)\right].
\label{sc}
\end{eqnarray}

With similar manipulations for quantum strong decoherence dynamics, we obtain
the quantum result
\begin{eqnarray}
\frac{\langle Q_{1}|\hat{\tilde{\rho}}(t)|Q_{2}\rangle}{\langle
Q_{1}|\hat{\tilde{\rho}}(0)|Q_{2}\rangle}=
\exp \left[i\phi_{q}(t)
-
\Delta Q^{2}\left( \frac{\Delta f(\overline{Q})}{\Delta Q}\right)^{2}B_{2}(t)\right],
\label{qresult2}
\end{eqnarray}
where $\phi_{q}(t)\equiv
\Delta Q f(\overline{Q})[\Delta f(\overline{Q})/\Delta
 Q]
 B_{1}(t)]/\hbar$,
 and
 \begin{eqnarray}
 S_{q}(t)=1-\int dQ_{1}dQ_{2}
 |\langle Q_{1}|\hat{\tilde{\rho}}(0)|Q_{2}\rangle|^{2}
 \exp\left[-2(\Delta Q)^{2}
 \left(\frac{\Delta f(\overline{Q})}{\Delta Q}\right)^{2}B_{2}(t)\right].
 \label{qc}
 \end{eqnarray}
 These results extend those in Ref. \cite{Haake} to nonlinear $f(Q)$ using
 a simple approach and demonstrate a direct classical analog to quantum
 strong decoherence dynamics.

Since $dB_{2}(t)/dt(t=0)=0$ and
$d^{2}B_{2}(t)/dt^{2}(t=0)=C_{b}/\hbar$, one finds that in the
short time limit, Eqs. (\ref{sc}) and (\ref{qc}) reduce to
previous perturbation results of $1/\tau_{c,1}$,
$1/\tau_{c,2}^{2}$, $1/\tau_{q,1}$, and $1/\tau_{q,2}^{2}$.
Furthermore, the classical results [Eqs. (\ref{cresult2}) and
(\ref{sc})] are again much similar to the quantum results [Eqs.
(\ref{qresult2}) and (\ref{qc})], with the only difference being
that $\Delta f/\Delta Q$  in the quantum expression is replaced by
$df/dQ$ in the classical result.

This result makes clear that our previous QCC results based upon
second-order perturbation theory are generalizable to all orders
of time in the strong decoherence case. In particular, defining
$\gamma_{c}(t)\equiv d\ln |\tilde{\rho}_{c}(Q_{1},Q_{2},t)|/dt$
and $\gamma_{q}(t)\equiv d\ln |\langle
Q_{1}|\hat{\tilde{\rho}}(t)|Q_{2}\rangle|/dt$, we have
$\gamma_{c}(t)=-(\Delta Q)^{2}
\left[df(\overline{Q})/d\overline{Q}\right]^{2}(dB_{2}(t)/dt)$,
and $\gamma_{q}(t) =-(\Delta Q)^{2}\left[\Delta
f(\overline{Q})/\Delta Q \right]^{2}(dB_{2}(t)/dt)$.
Then, in the case of linear and/or quadratic coupling, e.g.,
$f(Q)=aQ+bQ^{2}$, one has $\gamma_{c}(t)=\gamma_{q}(t)$ and
$S_{c}(t)=S_{q}(t)$, showing that there is perfect QCC in
decoherence dynamics for all times.

By contrast, in the case of nonlinear coupling, $\gamma_{c}(t)$ in
general does not saturate with increasing $\Delta Q$ whereas
$\gamma_{q}(t)$ does saturate for bounded $|f(Q)|$.  As such, in
the limit of large $\Delta Q$, one has $|\gamma_{c}(t)|
>>|\gamma_{q}(t)|$ and thus $[1-S_{c}(t)]<< [1-S_{q}(t)]$ as $t$
increases, with $|\gamma_{c}(t)/\gamma_{q}(t)|$ independent of $\hbar$.
This observation is of conceptual importance: it says that decoherence can
dramatically improve QCC, but as far as some detailed characteristics of
decoherence dynamics are concerned, decoherence itself does not
necessarily suffice to ensure that the dynamics of quantum entropy
production equals that of classical entropy production. That is, even in
the presence of strong decoherence, subtle quantum classical differences
may persist in some measures (e.g., $1/[1-S_{q}(t)]$ vs. $1/[1-S_{c}(t)]$)
for all finite times. 
Note, however,
the entropy measures such as $1/[1-S_{q}(t)]$ are not a quantum
mechanical observables and hence do not allow one to directly measure the subtle
difference between classical and quantum decoherence dynamics at later times.

Thus, from both the perturbation and strong decoherence results, we obtain that QCC
depends critically upon the initial quantum state and the nature of 
the system-environment coupling.
This result should have an impact on our current understanding of decoherence 
even when the role of the dynamics of the system is important.
For example,  it is worthwhile reexamining  
the relationship between classical Lyapunov exponents
and decoherence rates in classically chaotic systems, since previous
studies \cite{zureketc} only dealt with 
the case of linear system-environment coupling.

In conclusion, we have examined, using analogous measures, the decoherence
dynamics of an initial quantum state coupled to a bath that is subjected
to either classical or quantum dynamics. Within the framework of a
second-order perturbative treatment and a strong decoherence theory, we
have exposed the system-independent
conditions under which the quantum decoherence dynamics
is either well, or poorly, approximated by classical dynamics.
Further studies
are ongoing to assess QCC in cases beyond the short time and strong
decoherence approximations. Preliminary computational
results \cite{gong2} support the conclusions drawn herein.

This work was supported by the U.S. Office of Naval Research and the
Natural Sciences and Engineering Research Council of Canada.

\pagebreak


\begin{thebibliography}{100}
\bibitem{zurekreview} W.H. Zurek, quant-ph/0105127;
D. Giulini {\it et al}.,
{\it Decoherence and the Appearance of a Classical World
in Quantum Theory} (Springer, New York, 1996).
\bibitem{gong} S. Habib, K. Shizume, and W.H. Zurek, \prl{\bf 80}, 4361 (1998);
J. Gong and P. Brumer, \pre{\bf 60}, 1643 (1999).
\bibitem{raizen} For example, V. Milner {\it et al}.,  \pre{\bf 61}, 7223 (2000);
M.B. d'Arcy {\it et al}., \pre{\bf 64}, 056233 (2001).
\bibitem{wilkie} For example, J. Wilkie and P. Brumer, Phys. Rev. A {\bf 55}, 27 (1997);
Phys. Rev. A {\bf 55}, 43 (1997); C. Jaff\'{e}, S. Kanfer, and P. Brumer, \prl{\bf
54}, 8 (1985).
\bibitem{qcbook} M.A. Nielsen and I.L. Chuang, {\it Quantum Computation
and Quantum Information} (Cambridge University Press, Cambridge, 2000).
\bibitem{brumerbook}
M. Shapiro and P. Brumer, Adv. Atom. Mol. and Opt.
Phys., {\bf 42}, 287 (2000).
\bibitem{miller}
H. Wang {\it et al}.,
\jcp{\bf 114}, 2562 (2001);
F. Grossmann, \jcp{\bf 103}, 3696 (1995);
A.M.O. de Almeida, preprint quant-ph/0208094;
V.S. Batista and P. Brumer, Phys. Rev. Lett. {\bf 89}, 143201 (2002).
\bibitem{gong2} H. Han, J. Gong, and P. Brumer, to be published.
\bibitem{linear-entropy}
P.C. Lichtner and J.J. Griffin, \prl{\bf 37}, 1521 (1976);
W.H. Zurek, S. Habib and J.P. Paz, \prl{\bf  70}, 1187 (1993);
X-P. Jiang and P. Brumer, Chem. Phys. Lett. {\bf 208}, 179 (1993).
\bibitem{kim} J.I. Kim {\it et al}., \prl{\bf 77}, 207
(1996).
\bibitem{perturbation-approaches} L.M. Duan and G.C. Guo, \pra{\bf 56},
4466 (1997); D. Bacon, D.A. Lidar, and K.B. Whaley, \pra{\bf 60},
1944 (1999).
\bibitem{hu} B.L. Hu, J.P. Paz, and Y. Zhang, \prd{\bf 47}, 1576
(1993).
\bibitem{linear-models} R.P. Feynman and F.L. Vernon,
Ann. Phys. (N.Y.) {\bf 24}, 118 (1963);
A.O. Calderia and A.J. Leggett, Ann. Phys. (N.Y.) {\bf 149}, 374 (1983);
B.L. Hu, J.P. Paz, and Y. Zhang, \prd{\bf 45}, 2843 (1992).
\bibitem{footnote1} It is not surprising, in retrospect, that the
classical and quantum results are in perfect agreement in the
cases where the bath is harmonic, the coupling is linear or
quadratic, and the system can be neglected. In this case it is
expected that the quantum and classical Liouville dynamics are
identical. What is important, however, is that this equivalence
can be extended to the process of decoherence, a phenomenon assumed
to be solely quantum in nature.
\bibitem{saturation-note}C.C. Cheng and M.G. Raymer, \prl{\bf 82}, 4807 (1999);
J.R. Anglin, J.P. Paz,
and W.H. Zurek, \pra{\bf 55}, 4041 (1997).
\bibitem{Haake}D. Braun, F. Haake, and W.T. Strunz, \prl{\bf 86}, 2913 (2001).
\bibitem{zureketc}
W.H. Zurek and J. P. Paz, \prl{\bf 72}, 2508 (1994); A. K.
Pattanayak, Phys. Rev. Lett. {\bf 83}, 4526 (1999);  D. Monteoliva
and J.P. Paz, \prl{\bf 85}, 3373 (2000).

\end{thebibliography}
\end{document}